\documentstyle[12pt]{article}
\global\parskip 6pt

\normalsize

\let\oldtheequation=\theequation
\def\doteqs#1{\setcounter{equation}{0}
            \def\theequation{{#1}.\oldtheequation}}

\newcounter{sxn}
\def\sx#1{\addtocounter{sxn}{1} \bigskip\medskip \goodbreak
\noindent{\large\bf
\centerline{\thesxn.~~#1}} \nobreak \medskip}
\def\sxn#1{\sx{#1} \doteqs{\thesxn}}

\newcounter{axn}

\def\br{\begin{thebibliography}{abc}}
\def\er{\end{thebibliography}}

\def\be{\begin{equation}}
\def\ee{\end{equation}}
\def\bea{\begin{eqnarray}}
\def\eea{\end{eqnarray}}

\begin{document}
\begin{flushright}
\hfill{SINP-TNP/01-02}\\
\hfill{hep-th/0102051}
\end{flushright}
\vspace*{1cm}
\thispagestyle{empty}
\centerline{\large\bf Near-Horizon Conformal Structure of Black Holes}
\bigskip
\begin{center}
Danny Birmingham\footnote{Email: dannyb@pop3.ucd.ie}\\
{\em Department of Mathematical Physics\\
University College Dublin\\
Belfield, Dublin 4, Ireland}\\
\vspace*{.5cm}
Kumar S. Gupta\footnote{Email: gupta@tnp.saha.ernet.in}
and Siddhartha Sen\footnote{Email: sen@tnp.saha.ernet.in; On leave
from: School of Mathematics, Trinity College Dublin, Ireland}\\
{\em Saha Institute of Nuclear Physics\\
1/AF Bidhannagar\\
Calcutta - 700 064, India}\\
\end{center}
\vskip.5cm

\begin{abstract}
The near-horizon properties of a black hole are studied within an algebraic
framework, using a scalar field as a simple probe to analyze the geometry.
The operator $H$ governing the near-horizon dynamics of the scalar field
contains an inverse square interaction term. It is shown that
the operators appearing in
the corresponding algebraic description belong to the
representation space
of the Virasoro algebra. The operator $H$ is studied using the
representation theory of the Virasoro algebra. We observe
that the wave functions exhibit scaling behaviour in a band-like region
near the horizon of the black hole.
\end{abstract}
\vspace*{1cm}
\begin{center}
February 2001
\end{center}
\newpage

\sxn{Introduction} 

The relation between the physics of black holes and conformal field
theory has been explored recently in a variety of
contexts \cite{Strom, carlip, solo}.
In particular, the near-horizon symmetry structure of
general black holes in arbitrary dimensions
(including the Schwarzschild case) has been studied \cite{carlip, solo}.
By imposing suitable boundary conditions at the horizon, it was
shown that the relevant algebra of surface deformations contains
a Virasoro algebra in the $(r-t)$-plane. This analysis is based
on an extension of the Brown-Henneaux algebra of three-dimensional
anti-de Sitter gravity \cite{brown}.
In the latter case, purely classical considerations
lead to the existence of an asymptotic symmetry algebra containing
two copies of the Virasoro algebra.

In a separate line of development, it was found that the dynamics of
particles or scalar fields near the horizon of a black hole
is associated
with a Hamiltonian containing an inverse square 
potential \cite{town1,town2,trg}. Since the scalar field can be
viewed as a tool to probe the near-horizon geometry of the black hole,
its dynamics should reveal any underlying symmetry of the system.
Indeed, such a Hamiltonian was shown to have conformal symmetry 
quite some time ago \cite{fubini}, and this idea has
been further
explored recently \cite{vasiliev,town1,town2}.

In this paper, we provide a synthesis of the ideas
appearing in the above approaches within an algebraic framework.
Since our essential interest is in the near-horizon geometry of the
black hole, it is useful to restrict attention to a very simple
probe. Thus, we consider the time-independent modes
of a scalar field in the black hole background.
In particular, we study the case of a such a field
in the background of a Schwarzschild black hole.
The Klein-Gordon operator $H$ governing the
dynamics of the probe contains an inverse square potential term \cite{trg}.
It is shown that $H$ can be written
in a factorized form, which leads to an algebraic description
of the system in terms of the enveloping algebra of the Virasoro algebra.
The inverse square interaction term plays a
crucial role in obtaining this result. It may be noted 
that previous works in this
direction did not treat the interaction term algebraically.
Incorporating this term  within the algebraic framework leads to a 
structure that contains approximately half of all the Virasoro generators. 
The requirement of a unitary representation of the resulting algebra allows
us to include the remaining generators.
We then describe the spectrum of $H$ in terms of the
wedge representations of the Virasoro algebra \cite{kac}.

It should be noted that the
inverse square term also plays an important role in determining the
self-adjoint extensions of the Klein-Gordon operator \cite{trg}.
In general, the corresponding wave functions
violate scaling in the near-horizon region.
However, we show that scaling behaviour is present in a small band-like
region near the horizon, for certain choices of the self-adjoint extension.
These self-adjoint extensions
thus play a crucial role in providing a consistent picture of the whole
analysis. The existence of this band region is reminiscent
of the stretched horizon picture of black holes \cite{suss}.

This paper is organized as follows. In section 2, we study the
example of a scalar field probing the near-horizon properties of the
Schwarzschild black hole. The operator governing the
dynamics of the time-independent modes is written in a factorized form.
It is shown that the resulting factors lead to an algebraic
description in terms of the enveloping algebra of the Virasoro generators.
Section 3 discusses the properties of this algebra 
in terms of the wedge representations of Ref. \cite{kac}. This leads
to an algebraic description of the spectrum of the time-independent
Klein-Gordon operator.
The near-horizon scaling behaviour at the quantum level is discussed in
section 4. We conclude in section 5 with a brief discussion
regarding the application of our results to more general
black holes.

\sxn{Algebraic Formulation of the Near-Horizon Dynamics}

In this section, we consider the case of a scalar field probing the
near-horizon geometry of a Schwarzschild black hole.
We shall restrict the analysis to the
time-independent modes of the scalar field. The Klein-Gordon
operator governing the
near-horizon dynamics can then be written as \cite{trg}
\begin{equation}
H = - {\frac{d^2}{dx^2}} + {a \over x^2},
\end{equation}
where $a$ is a real dimensionless constant, and $x \in [0, \infty]$ is the
near-horizon coordinate. For the Schwarzschild background,
we have $a = - \frac{1}{4}$.
For the moment, however, we can consider a general value of $a$.

The essential point to note is that the operator $H$ can be factorized
as
\be
H = A_+A_-,
\label{H}
\ee  
where
\be
A_{\pm} = \pm \frac{d}{dx} + \frac{b}{x},
\ee
and 
\be
b =  \frac{1}{2} \pm \frac{\sqrt{1 + 4a}}{2}.
\ee 
We note that $a = - \frac{1}{4}$ is the minimum value of $a$
for which $b$ is
real. For real values of $b$, $A_+$ and $A_-$ are formal
adjoints of each other (with 
respect to the measure $dx$), and consequently $H$ is formally
a positive quantity
(there are some subtleties to this argument arising from the
self-adjoint extensions of $H$ which will be discussed later).
When $a < - \frac{1}{4}$, $b$ is no longer real and
$A_+$ and $A_-$ are not
even formal adjoints of each other. However,
$H$ can still be factorized as in Eqn. (\ref{H}),
but it is no longer a positive definite quantity.
It can still be
made self-adjoint \cite{case}, but remains unbounded
from below; this case has been analyzed in \cite{ksg}.

Let us now define the operators
\bea
L_n &=& -x^{n + 1} \frac{d}{dx},~~~~~~~~~n \in {\mathbf Z},\\
P_m &=& \frac{1}{x^m},~~~~~~~~~m \in {\mathbf Z}.
\eea
In terms of these operators, $A_{\pm}$ and  $H$ can be written as
\bea
A_{\pm} &=& \mp L_{-1} + b P_1,\\
H &=& (- L_{-1} + b P_1) (L_{-1} + b P_1).
\eea
Thus,
$L_{-1}$ and $P_1$ are the  basic operators appearing in the factorization
of $H$. Taking all possible commutators of these operators between
themselves and with $H$, we obtain the following relations
\begin{eqnarray} {}
[ P_{m}, P_{n} ] &=& 0,\\ {}
[ L_{m}, P_{n} ] &=& n P_{n - m},\\ {}
[ L_{m}, L_{n} ] &=& ( m - n) L_{m + n} + \frac{c}{12}
(m^{3} - m)\delta_{m+n, 0},\\ {}
[P_m, H] &=& m (m+1) P_{m + 2} + 2 m L_{-m -2}, \\{}
[L_m, H] &=& 2b(b-1)P_{2-m} - (m+1)(L_{-1}L_{m-1} + L_{m-1}L_{-1}).
\end{eqnarray}
Eqn. (2.11) describes a Virasoro algebra with central charge $c$.
Note that the algebra of the generators defined  in Eqn. (2.5)
would lead to
$[ L_{m}, L_{n} ] = ( m - n) L_{m + n}$. However, this algebra
is known to
admit a non-trivial central extension. Moreover, for any
irreducible unitary highest weight representation of
this algebra,
$c \neq 0$ \cite{godd}. For these reasons,
we have included the central term explicitly
in Eqn. (2.11).

Eqns. (2.9 - 2.11) describe the semidirect product of the Virasoro algebra
with an abelian algebra satisfied by the shift operators $\{P_{m}\}$.
Henceforth, we denote this semidirect product algebra by
${\cal {M}}$. 
Note that $L_{- 1}$ and $P_1$ are the only  generators that appear in $H$.
Starting with these two generators, and using Eqns. (2.12) and (2.13),
we see that the only operators which appear are the Virasoro generators
with negative index (except $L_{-2}$), and the shift
generators with positive index.
Thus, $L_m$ with $m \geq 0$ and $P_m$
with $m \leq 0$ do not appear in the above expressions. In the next
section, we will discuss how these quantities are generated.

Although the  algebra of Virasoro and shift generators
has a semidirect product structure, the operator $H$
however does not belong to this algebra.
This is due to the fact that the right-hand side of Eqn. (2.13)
contains products of the Virasoro generators. While
such products are not
elements of the algebra, they do
belong to the corresponding enveloping algebra.
The given system is thus seen to be described by the 
enveloping algebra of the Virasoro generators, together with the abelian
algebra of the shift operators. This algebraic system has been extensively
studied in the literature \cite{kac}.

\sxn {Representation}

We wish to discuss the representation theory of the algebra
${\cal{M}}$, and the implications for the quantum properties of the
Klein-Gordon operator $H$.
The eigenvalue equation of interest is
\be
H | \psi \rangle =  E | \psi \rangle,
\ee
with the boundary condition that $\psi (0) = 0$. We are especially
interested in the bound state sector of $H$.  
As we have seen, the operator $H$ can be expressed in terms
of certain operators that belong to the
algebra ${\cal {M}}$. This observation allows us to
give a description of the  states of $H$ in terms of the
representation spaces of ${\cal {M}}$.
We first recall the relevant aspects of the representation theory
of ${\cal{M}}$.

Following \cite{kac},
we introduce the space $V_{\alpha, \beta}$ of densities
containing elements of the form
$P(x) x^\alpha (dx)^\beta$, where $\alpha, \beta$ are complex
numbers, in general.
Here, $P(x)$ is an arbitrary polynomial in $x$ and $x^{-1}$,
where $x$  is now treated as a complex variable.
It may be noted that the algebra ${\cal {M}}$ remains unchanged even
when $x$ is complex. It is known that
$V_{\alpha, \beta}$ carries a representation of the algebra ${\cal {M}}$.
The space $V_{\alpha, \beta}$ is spanned by a set of vectors,
$\omega_m = x^{m + \alpha} (dx)^\beta$, where $m \in {\mathbf Z}$.
The Virasoro generators and the shift operators have the following
action
on the basis vectors $\omega_m$,
\bea
P_n (\omega_m) &=& \omega_{m-n},\\
L_n (\omega_m) &=& - (m + \alpha + \beta + n \beta) \omega_{n + m}.
\eea 
The representation $V_{\alpha, \beta}$ is reducible
if  $\alpha \in {\mathbf Z}$ and if $\beta = 0$ or $1$;
otherwise it is irreducible.

The requirement of unitarity of the representation $V_{\alpha, \beta}$
leads to several important consequences. In any unitary representation of 
${\cal {M}}$, the Virasoro generators must satisfy the condition
${L_{- m}^{\dagger}} = L_m$. In the previous section, we saw
that
$L_{-2}$ and $L_m$ for $m \geq 0$ did not appear in the algebraic
structure generated by the basic operators appearing in
the factorization of $H$.
However, the requirement of a unitary
representation now leads to the inclusion of $L_m$ for
$m > 0$. The remaining generators now appear through
appropriate commutators, thus completing the algebra ${\cal M}$.

Unitarity also constrains the parameters $\alpha$ and
$\beta$, which must now satisfy the conditions
\bea
\beta + \bar{\beta} &=& 1,\\
\alpha + \beta &=& {\bar{\alpha}} + {\bar{\beta}},
\eea
where ${\bar{\alpha}}$ denotes the complex conjugate of $\alpha$.
Finally, the central charge $c$ in the representation $V_{\alpha, \beta}$
is given by
\be
c (\beta) = -12 \beta^2 + 12 \beta -2.
\ee

The above representation of ${\cal {M}}$ can now be used to analyze the
eigenvalue problem of Eqn. (3.1). We would like to have a series solution to
the differential Eqn. (3.1), and consequently choose
an ansatz for the  wave function $\mid\!\psi \rangle$ 
given by
\be
| \psi \rangle = \sum^{\infty}_{n=0} c_n \omega_n.
\ee     
Furthermore, the
operator $H$, as written in Eqn. (\ref{H}), has a well-defined action on
$| \psi \rangle $. From Eqn. (3.3), it may be seen that
\be
L_{-1} (\omega_n) = - (n + \alpha) \omega_{n-1},
\ee
which is independent of $\beta$. Therefore, it appears that an
eigenfunction of $H$ may be constructed from elements of
$V_{\alpha, \beta}$
for arbitrary  $\beta$. However, the unitarity conditions of
Eqns. (3.4-3.5) put
severe restrictions on $\beta$, as we shall see below.

The indicial equation 
obtained by substituting Eqn. (3.7) in Eqn. (3.1) gives
\be
\alpha = b,\;{\mathrm{or}}~(1 - b).
\ee
To proceed, we analyze the cases (i) $a \geq - \frac{1}{4}$,
and (ii) $ a < - \frac{1}{4}$ separately.

\noindent
(i) $a \geq - \frac{1}{4}$.

This is the main case of interest as it includes
the value of $a$ for the
Schwarzschild background.
It follows from Eqn. (2.4) and
Eqn. (3.9) that $b$ and $\alpha$ are real. The unitarity condition
of Eqn (3.4) now fixes the value of $\beta = \frac{1}{2}$, and the
corresponding central charge is given by $c = 1$. It may be noted that 
relation of the central charge calculated here to that appearing in
the calculation of black hole entropy depends on geometric properties
of the black hole in question. We do not address this issue here.
Thus, we see that for the Schwarzschild black hole, we have identified
the relevant representation space as $V_{1/2,1/2}$.

\noindent
(ii) $ a < - \frac{1}{4}$.

In this case, we can write $ a = - \frac{1}{4} - \mu^2$
where $\mu \in {\mathbf R}$.
It follows from Eqn. (2.4), that $b = \frac{1}{2} \pm i \mu$. Eqn. (3.9)
then gives $\alpha = \frac{1}{2} \pm i \mu$, or $-\frac{1}{2} \mp i \mu$.
Let us take the   
case when $\alpha = \frac{1}{2} + i \mu$, the other
cases being similar. From Eqns. (3.4) and (3.5), we find
$\beta = \frac{1}{2} - i \mu$.
The value of the corresponding central charge is given by 
$c = 1 + 12 \mu^2$. The operator $H$ in this case can be made self-adjoint
but its spectrum remains unbounded from below \cite{case,ksg}. The algebraic
description, however, always leads to a well-defined representation.

We return now to the eigenvalue problem for the
differential operator $H$, and focus attention on the Schwarzschild
background, for which $a = -\frac{1}{4}$.
As already mentioned, we are interested in the
bound state sector of $H$.
These states have negative energy and satisfy
Eqn. (3.1) with energy $-E$, where $E>0$.
The wave function satisfies the boundary condition
$\psi (0) = 0$.
This may seem contradictory to the statement in section 2,
which claimed that $H$ as written in Eqn. (2.2) is a positive quantity.
The resolution of this apparent paradox is as follows.
It is known that the operator $H$ admits a
one-parameter family of self-adjoint extensions labelled
by a $U(1)$
parameter $e^{iz}$, where $z$ is real \cite{narn,reed,trg}.
For $ a = - \frac{1}{4}$, there is an
infinite number of bound states
for a given self-adjoint extension $z$. 
In all these cases, $A_{+}$ and $A_{-}$ are not adjoints of each other,
and consequently $H$ is not a positive quantity.
The eigenfunctions and eigenvalues of $H$ in this case are given by
\cite{trg}
\bea
\psi_n (x) &=& N_n \sqrt{x} K_0\left( \sqrt{E_n} x\right),\\
E_n &=& {\exp}\left[\frac{\pi}{2} (1 - 8n) {\cot} \frac{z}{2}\right],
\eea
where $n$ is an integer, $N_n$ is a normalization factor,
and $K_{0}$ is the modified Bessel function.

We have thus shown how to obtain the spectrum of $H$ using the
representation of the algebra ${\cal {M}}$.
In the next section, we shall
analyze the properties of this spectrum in the near-horizon region.

\sxn{Scaling Properties}

As we have seen, the Virasoro algebra
plays an important role in determining the spectrum of $H$.
Since this operator is associated with a probe of the near-horizon
geometry, one might expect that the corresponding wave functions
would exhibit certain scaling behaviour in this
region.

Firstly,  let us recall that
the horizon in this picture is located at $x = 0$. However,
the wave functions
$\psi_n$ vanish at $x = 0$, and therefore do not exhibit
any non-trivial scaling.
Nevertheless, it is of interest to examine the behaviour of
the wave functions near the horizon. For $x \sim 0$,
the wave functions have the form
\be
\psi_n = N_n \sqrt{x}\left(A - \ln \left( \sqrt{E_n} x\right)\right),
\ee
where $A = \ln 2 - \gamma$, and $\gamma$ is Euler's constant \cite{abr}.
While the logarithmic term, in general, breaks the scaling property,
one notices that it vanishes at the  point
$x_0 \sim 1/\sqrt{E_n}$,
where the wave functions exhibit a scaling
behaviour. 
The entire analysis so far, including the existence of the Virasoro
algebra, is valid only in the near-horizon region of the black hole.
Therefore, consistency of the above scaling behaviour
requires that $x_0$  belongs to the near-horizon region.
The minimum value of $x_0$ is obtained when $E_n$ is maximum. When the
parameter $z$ appearing in the self-adjoint extension of $H$
is positive,
the maximum value of $E_n$ is given by 
\be 
E_0 = {\exp}\left[\frac{\pi}{2} {\cot} \frac{z}{2}\right].
\ee
However, when $z$ is negative, the maximum value of $E_n$ is obtained when
$n \rightarrow \infty$. In this case, $x_0 \rightarrow 0$ where, as we have
seen before, the wave function vanishes and scaling becomes trivial. We
therefore conclude that 
\be
x_0 \sim \frac{1}{\sqrt{E_0}},~ z > 0
\ee
is the minimum value of $x_0$.
It remains to show that $x_0$ given by Eqn. (4.3) belongs to
the near-horizon region. We first note that we are free
to set $z$ to an arbitrary positive value.
Thus, we consider  $z >0 $ such that
${\cot} \frac{z}{2} >> 1$; this is achieved by choosing
$z \sim 0$.
For all such $z$,  we find that $x_0$ is small but nonzero, and
thus belongs to the near-horizon region.
In effect, we can use the freedom in the choice of
$z$ to restrict $x_0$ to the near-horizon region.

We now consider a band-like region
$\Delta = [x_{0} -\delta/\sqrt{E_{0}}, x_{0} + \delta/\sqrt{E_{0}}]$,
where $\delta \sim 0$ is real and positive. The region $\Delta$ thus belongs
to the near-horizon region of the black hole.
At a point $x$ in the region $\Delta$,
the leading behaviour of $\psi_n$ is given by 
\be
\psi_n = N_n \sqrt{x} \left(A + 2 \pi n\; {\cot} \frac{z}{2}\right).
\ee
Thus, all the eigenfunctions of $H$ exhibit a scaling behaviour,
i.e. $\psi_n \sim \sqrt{x}$, in the near-horizon region $\Delta$.
It should be stressed that this analysis is made possible
by utilizing the freedom in the choice of $z$.
The parameter $z$, which labels the self-adjoint extensions
of $H$, thus plays a crucial role in
establishing the self-consistency of this analysis.

We conclude this section with the following remarks:

\noindent
1. A particular choice of $z$ is
equivalent to a choice of domain for the differential operator $H$.
Physically,
the domain of an operator is specified by boundary conditions. A specific
value of $z$ is thus directly related to a specific choice of boundary
conditions for $H$. Thus,
we see that the system exhibits non-trivial scaling behaviour
only for a certain class of boundary
conditions.
These boundary conditions play a conceptually
similar role to the fall-off 
conditions as discussed in Ref. \cite{carlip,solo}.

\noindent
2. The analysis above provides a qualitative argument which suggests that
the scaling behaviour in the presence of a black hole
should be observed within a region
$\Delta$. Although $\Delta$ belongs to the near-horizon region
of the black hole, it
does not actually contain the event horizon. Our picture is thus similar in
spirit to the stretched horizon scenario of Ref. \cite{suss}.
\newpage

\sxn{Conclusion}

In this paper, we have analyzed the near-horizon properties of the
Schwarzschild black hole, using a  scalar field as a
simple probe of the system. We restricted attention
to the time-independent modes of the scalar
field, and this allowed us
to obtain a number of interesting results
regarding the near-horizon properties of the black hole.
It is possible
that more sophisticated probes of general field
configurations may lead to additional information.

The factorization of $H$,
leading to the algebraic formulation of section 2,
is a process which appears to be essentially classical.
However, the central charge in
the algebra ${\cal {M}}$ goes beyond the classical framework, as it
arises from the requirement of a non-trivial representation.
As discussed, the algebra appearing in Eqns. (2.9-2.13) does not at
first contain all the Virasoro generators.
The requirement of unitarity of the
representation leads to the inclusion of all the generators.
It is thus fair
to say that the full Virasoro algebra appears in our
framework only at the
quantum level. The operator $H$ does not belong to ${\cal {M}}$ but 
is contained in the enveloping algebra of the Virasoro generators.  
The enveloping algebra is the natural tool that 
is used to obtain representations of ${\cal {M}}$. Thus, even
though $H$ is not an element of ${\cal {M}}$, it nevertheless
has a well-defined
action in any representation of ${\cal {M}}$. It is this feature
that makes the algebraic description useful.

In section 3,
we summarized some results from the representation theory of
${\cal {M}}$. The operator $H$ is now treated at the quantum level,
and the corresponding eigenvalue problem is studied
using the representations
of ${\cal {M}}$. Unitarity again plays a role in restricting the space of
allowed representations. It is interesting to note that for all values of
the coupling $a \geq - \frac{1}{4}$, the value of the central charge in the
representation of ${\cal {M}}$ is equal to 1. Other black holes
which have $a$ in this range would exhibit a universality in this
regard. As mentioned in section 3, the relationship of $c$ calculated here
to that appearing in the entropy calculation of a particular black hole
would depend
on other factors which are likely to break the universality.

If a Virasoro algebra is associated with the near-horizon dynamics,
then some reflection of it should appear in the spectrum of $H$. In
particular, we can expect that the wave functions of $H$ in the
near-horizon region should exhibit scaling behaviour.
Such a property was indeed
found in a band-like region near the horizon.
It is interesting to note that
this band excludes the actual horizon. This is similar in spirit to the
stretched horizon scenario of black hole dynamics.
The parameter $z$ describing the
self-adjoint extensions of $H$ is restricted to a set of values in this
process. This implies that the near-horizon wave functions
exhibit scaling behaviour only for a certain class of boundary
conditions.
It is important to note that boundary
conditions also played a crucial role in proving the existence of a Virasoro
algebra in Ref. \cite{carlip, solo}.
This feature provides a common thread in these
different approaches towards the problem.

It is known that the near-horizon dynamics of various black holes
is described by an operator of the form $H$
\cite{trg,town1}, for different values of $a \geq -\frac{1}{4}$.
Any such operator can be factorised as
in Eqn. (2.2) and the above analysis will also apply
to these black holes.
It has been claimed
in \cite{carlip, solo} that a Virasoro algebra is associated with a
large class of black holes in arbitrary dimensions.
It seems plausible that the near-horizon dynamics
of probes in the background of these black holes
would be described by an operator of the form of $H$.

\newpage
\noindent
{\bf Acknowledgements}

K.S.G. would like to thank A.P.Balachandran for discussions.

\bibliographystyle{unsrt}

\end{document}